\documentclass[letterpaper, 10 pt, conference]{ieeeconf}  

\IEEEoverridecommandlockouts                              

\usepackage{balance}
\usepackage{mathrsfs}

\usepackage{epsfig} 
\usepackage{mathptmx} 
\usepackage{times} 
\usepackage{amsmath} 
\usepackage{amssymb}  
\usepackage{algorithm}
\usepackage{algorithmic}
\usepackage{accents}

\makeatletter
\DeclareRobustCommand{\cev}[1]{%
  \mathpalette\do@cev{#1}%
}
\newcommand{\do@cev}[2]{%
  \fix@cev{#1}{+}%
  \reflectbox{$\m@th#1\vec{\reflectbox{$\fix@cev{#1}{-}\m@th#1#2\fix@cev{#1}{+}$}}$}%
  \fix@cev{#1}{-}%
}
\newcommand{\fix@cev}[2]{%
  \ifx#1\displaystyle
    \mkern#23mu
  \else
    \ifx#1\textstyle
      \mkern#23mu
    \else
      \ifx#1\scriptstyle
        \mkern#22mu
      \else
        \mkern#22mu
      \fi
    \fi
  \fi
}

\makeatother

\newcommand{\qed}{\hfill\square}

\newtheorem{theorem}{Theorem}
\newtheorem{lemma}{Lemma}

\newtheorem{definition}{Definition}

\usepackage{graphicx}
\graphicspath{{./figures}}

\title{\LARGE \bf
Modularized Control Synthesis for Complex Signal Temporal Logic Specifications
}

\newcommand{\F}{\mathsf{F}}
\newcommand{\G}{\mathsf{G}}
\newcommand{\useq}{\mathbf{u}}
\newcommand{\xseq}{\mathbf{x}}

\author{Zengjie Zhang and
        Sofie Haesaert 
\thanks{This work was supported by the European project SymAware under the grant Nr. 101070802.}
\thanks{Zengjie Zhang and Sofie Haesaert are with the Department of Electrical Engineering, Eindhoven University of Technology,
        PO Box 513, 5600 MB Eindhoven, Netherlands.
        {\tt\small \{z.zhang3, s.haesaert\}@tue.nl}}%
}

\begin{document}

\newcommand{\Ta}{\mathcal{T}}
\newcommand{\xt}[1]{x^{#1}}
\maketitle
\thispagestyle{empty}
\pagestyle{empty}

\begin{abstract}

The control synthesis of a dynamic system subject to a signal temporal logic (STL) specification is commonly formulated as a mixed-integer linear/convex programming (MILP/ MICP) problem. Solving such a problem is computationally expensive when the specification is long and complex. In this paper, we propose a framework to transform a long and complex specification into separate forms in time, to be more specific, the logical combination of a series of short and simple subformulas with non-overlapping timing intervals. In this way, one can easily modularize the synthesis of a long specification by solving its short subformulas, which improves the efficiency of the control problem. We first propose a syntactic timing separation form for a type of complex specifications based on a group of separation principles. Then, we further propose a complete specification split form with subformulas completely separated in time. Based on this, we develop a modularized synthesis algorithm that ensures the soundness of the solution to the original synthesis problem. The efficacy of the methods is validated with a robot monitoring case study in simulation. Our work is promising to promote the efficiency of control synthesis for systems with complicated specifications.

\end{abstract}

\section{INTRODUCTION}

Signal temporal logic (STL) is widely used to specify requirements for robot systems~\cite{plaku2016motion, li2021vehicle}, due to its advantage in specifying real-valued signals with finite timing bounds~\cite{raman2014model}. 
System control with STL specifications renders a synthesis problem that can be solved by mixed integer linear/convex programming (MILP/MICP)~\cite{raman2014model, kurtz2021more}. Based on this a closed-loop controller can be developed using model predictive control (MPC)~\cite{lindemann2021reactive, salamati2021data}.
However, solving a MILP/MICP problem is computationally expensive and time-consuming, especially for complex STL formulas with long timing intervals since the computational load grows drastically as the number of the integer variables increases (exponentially in the worst case)~\cite{kurtz2022mixed}. Thus, computational complexity has become a bottleneck of the control synthesis of complex STL specifications, especially those with time-variant specifications~\cite{srinivasan2020control} and fixed-order constraints~\cite{lindemann2018control}. One effective approach is the model-checking-based method which transforms an STL formula into an automaton with strict timing bounds~\cite{gundana2021event}. This method is usually less complex than an optimization problem since it is only concerned with a feasible solution. Control barrier functions (CBF)~\cite{lindemann2018control} and funnel functions~\cite{liu2022compositional} are also used to simplify the STL synthesis problems.

Another direction of reducing the complexity is to decompose a long and complex STL formula into several shorter and simpler subformulas and solve them sequentially. A subformula refers to a simple STL formula that serves as a primitive unit of a complex formula~\cite{alatartsev2015robotic}. This indicates the possibility of splitting a big planning problem into several smaller problems and solving them one by one in the order of time, which forms the essential thought of modularized synthesis. This idea is straightforward from a practical perspective: a complex task is usually composed of a series of smaller subtasks that have independent objectives and are ordered in time. For example, a typical food delivery task includes three subtasks: picking up the order at the restaurant, navigating to the customer, and performing the delivery. Finishing these subtasks means accomplishing the overall task. The advantage of this approach is based on the assumption that solving a subtask may be substantially simpler than directly solving the original overall task. 

However, modularized synthesis based on specification decomposition is not trivial and brings up two major challenges. Firstly, the decomposed specification has to ensure \textit{soundness}, i.e., any feasible solution of the modularized synthesis must also be a feasible solution of the original specification. This is important to ensure the efficacy of the specification decomposition and modularized synthesis~\cite{wongpiromsarn2012receding}. Secondly, the subformulas may have overlapping timing intervals which indicate the dependence coupling among these subformulas. In this case, each subformula should not be synthesized independently but should incorporate the coupling with its overlapping subformulas. In existing work, the soundness of specification separation is ensured by \textit{syntactic separation} as partially discussed in~\cite{hunter2013expressive, bae2019bounded}. Recently, model checking based on specification decomposition has been studied for a fragment of STL formulas~\cite{yu2022model}. Nevertheless, the coupling issues among subformulas have not been well resolved by the existing work. To our knowledge, there is no other existing work discussing the modularized synthesis of STL formulas, although we believe it to be a promising technology for the efficient synthesis of complex specifications.

In this paper, we investigate the modularized synthesis of complex STL specifications based on timing separation. We specifically look into a fragment of STL formulas composed of complex temporal operators for which interval overlapping can not be resolved by purely using syntactic separation. Besides proposing several complementary syntactic separation principles to the existing work~\cite{hunter2013expressive, bae2019bounded}, we also provide a sufficient separation method for this STL fragment with the overlapping between subformulas eliminated. In such a way, we develop a modularized synthesis algorithm for the separated specification by transforming the overall synthesis problem into several small planning problems with reduced complexity, achieving higher efficiency than directly solving the original problem. The main contributions are as follows. 

1). Proposing a syntactic timing separation form of a fragment of STL formulas that is proven to be syntactically equivalent to the original specification.

2). Proposing a complete splitting form of this STL fragment which is proven to be sound in semantics.

3). Developing a modularized synthesis algorithm for the complete splitting form, which ensures soundness but less complexity than the original specification synthesis problem.

The rest of the paper is organized as follows. Sec.~\ref{sec:prel} introduces the preliminary knowledge of this paper. In Sec.~\ref{sec:main}, we present our main results on specification separation and modularized synthesis. Sec.~\ref{sec:casestudy} provides a simulation case study to validate the efficacy of the proposed modularized synthesis method. Finally, Sec.~\ref{sec:conclu} concludes this paper.

\textit{Notations:} We use $\mathbb{R}$ and $\mathbb{R}^n$ to denote the sets of real scalars and $n$-dimensional real vectors. We also use $\mathbb{N}$ and $\mathbb{N}^+$ to denote natural numbers and positive natural numbers.

\textit{Proofs:} All proofs of this paper are in the Appendix.

\section{Preliminaries and Problem Statement}\label{sec:prel}

\subsection{Signal Temporal Logic (STL)}\label{sec:stl}

Specifications in Signal Temporal Logic (STL) can quantify requirements on real-valued signals. In this paper, we are concerned with discrete-time signals $\xseq_{[\,0,\,L\,]}\!:=\!x_0x_1\cdots x_{L}$, where $L\!\in\!\mathbb{N}^+$ denotes the length of the signal and $x_k\!\in\! \mathbb{R}^n$ is the value of the signal at time $k\!\in\!\{0,1,\cdots,L\}$. 
With  $\xseq_{[\,k_1,k_2\,]}:=x_{k_1} x_{k_1+1}\ldots x_{k_2}$, we denote a segment of $\xseq$ or equivalently $\xseq_{[\,0,\,L\,]}$ with timing points $0\!\leq\!k_1\!\leq\!k_2\!\leq\!L$.
The syntax of STL is recursively defined as
\begin{equation}\label{eq:stldef}
\varphi ::= \top \,|\, \mu \,|\, \lnot \varphi \,|\, \varphi_1 \wedge \varphi_2 \,|\, \varphi_1 \mathsf{U}_{[\,a,\,b\,]} \varphi_2,
\end{equation}
where $\varphi_1, \varphi_2$ are STL formulas, $\lnot$, $\wedge$ are operators \textit{negation} and \textit{conjunction}, $\mu$ is a predicate that evaluates a predicate function $\eta: \mathbb{R}^n \rightarrow \{\, \top,\,\bot\,\}$ by $\displaystyle \mu = \left\{ \begin{array}{ll}
\top & \mathrm{if} ~ \eta(x_k) \geq 0 \\
\bot & \mathrm{if} ~ \eta(x_k) < 0
\end{array} \right.$, for discrete time $k$, and $\mathsf{U}_{[\,a,\,b\,]}$ is the \textit{until} operator bounded with time interval $[\,a,\,b\,]$, where $a,b \in \mathbb{N}$ and $a \leq b$. 

The semantics of STL are given as follows.
 We denote the satisfaction of $\varphi$ at time $k$  by $\xseq$ as $(\xseq, k) \vDash \varphi$. Furthermore, we have that
$(\xseq, k) \vDash \mu \leftrightarrow \eta(x_k) \geq 0$;
$(\xseq, k) \vDash \lnot \varphi \leftrightarrow \lnot((\xseq, k) \vDash \varphi)$;
$(\xseq, k) \vDash \varphi_1 \wedge \varphi_2 \leftrightarrow (\xseq, k) \vDash \varphi_1$ and $(\xseq,k) \vDash \phi_2$;
$(\xseq, k) \vDash \varphi_1 \mathsf{U}_{[\,a,\,b\,]} \varphi_2$ $\leftrightarrow$ $\exists \, k' \in [\,k+a,\,k+b\,]$, such that $(x, k') \vDash \varphi_2$, and $(x,k'') \vDash \varphi_1$ holds for all $k'' \in [\,k, \,k'\,]$.
Besides, additional operators \textit{disjunction}, \textit{eventually}, and \textit{always} are, respectively, defined as $\varphi_1 \vee \varphi_2 = \lnot \left( \lnot \varphi_1 \wedge \lnot \varphi_2 \right)$, $\F_{[\,a,\,b\,]} \varphi = \top \mathsf{U}_{[\,a,\,b\,]} \varphi$, and $\G_{[\,a,\,b\,]} \varphi = \lnot \F_{[\,a,\,b\,]} \lnot \varphi$. When $k\!=\!0$, we also write $(\xseq,0) \vDash \varphi$ as $\xseq \vDash \varphi$.
The \textit{length}~\cite{maler2004monitoring}
of an STL, $\mathcal{L}(\varphi)$, is recursively defined as $\mathcal{L}(\mu) = 0$, $\mathcal{L}(\lnot \varphi) = \mathcal{L}(\varphi)$, $\mathcal{L}(\varphi_1 \wedge \varphi_2) = \max\{ \mathcal{L}(\varphi_1), \mathcal{L}(\varphi_2) \}$, $\mathcal{L}(\varphi_1 \mathsf{U}_{[\,a,\,b\,]} \varphi_2) = b + \max\{ \mathcal{L}(\varphi_1), \mathcal{L}(\varphi_2) \}$, which represents the maximum time it takes to determine the truth of the formula $\varphi$.

\subsection{Optimization-Based Specification Synthesis}\label{sec:sswss}

STL formulas are used to specify the requirements of a signal of a dynamic system. In this paper, we consider the following discrete-time dynamic system, 
\begin{equation}\label{eq:opti_sys}
x_{k+1} = f(x_k,\, u_k),
\end{equation}
where $x_k \in \mathbb{R}^n$ and $u_k \in \mathbb{U}$ are the state and the control input of the system at time $k$, where $\mathbb{U} \subseteq \mathbb{R}^m$ is the admissible control set, and $f:\mathbb{R}^n \times \mathbb{U} \rightarrow \mathbb{R}^n$ is a smooth vector field. Then, the control problem of the system can be formulated as the following optimization problem,
\begin{subequations}\label{eq:opti_sto}
\begin{alignat}{4}
&\textstyle \min_{\useq} \left(\sum_{k=0}^{L-1} u_k^{\!\top\!}u_k - \rho(\xseq, \varphi) \right) \label{eq:opti_max} \\
\mathrm{s.t.}~& ~\mathrm{eq}.~\eqref{eq:opti_sys}~\mathrm{and}~u_{k} \in \mathbb{U},~\forall\,k \in \{0, 1, \cdots, L-1\}, \label{eq:opti_hor}
\end{alignat}
\end{subequations}
where $L\!\in\!\mathbb{N}^+$ is the control horizon, $\useq \!=\!u_{0}u_{1}$ $\cdots$ $u_{L-1}$ and $\xseq \!=\!x_{0} x_{1} \cdots x_L$ are the \textit{open-loop} control and state signals, $\varphi$ is an STL formula with $\mathcal{L}(\varphi)\!=\!L$ to specify the requirements on the state signal $\xseq$, and $\rho(\xseq, \varphi)$ is the robustness of the satisfaction as defined in~\cite{nenzi2015specifying, fainekos2009robustness}, with $\varrho(\xseq, \varphi)\!>\!0 \!\leftrightarrow\! \xseq \!\vDash\! \varphi$. 
Eq.~\eqref{eq:opti_sto} renders an \textit{open-loop} control problem and can be solved using MICP~\cite{kurtz2022mixed} with an input interface $(x_0,\, L,\, \varphi)$.  

\subsection{Problem Statement}\label{sec:prob_state}

Solving problem~\eqref{eq:opti_sto} using MICP usually introduces heavy computational load due to the large number of integer variables brought up by the logical and temporal operators in the specification $\varphi$~\cite{kurtz2022mixed}. Usually, longer formulas introduce substantially more integer variables than shorter ones. In the worst case, a specification $\varphi$ may contain $N\in\mathbb{N}^+$ $\mathsf{G}$ or $\mathsf{F}$ subformulas with the same length $L\!=\!\mathcal{L}(\varphi)$. This requires $NL$ binary variables to determine the logical satisfaction of the complete specification, which leads to an exponential complexity $O(2^{NL})$. Therefore, the computational complexity of the synthesis problem is greatly dependent on the number and the lengths of subformulas.

In this paper, we intend to reduce the complexity of a synthesis problem~\eqref{eq:opti_sto} by separating a long specification $\varphi$ into shorter subformulas which can be solved by smaller optimization problems. 
Examples of such separation include the following principle for an until operator $\varphi \!=\! \varphi_1 \mathsf{U}_{(\,a,\,b\,)} \varphi_2$ with a separating point $\kappa \!\in\! \mathbb{N}^+$, $a \!<\! \kappa \!<\! b$~\cite{hunter2013expressive},
\begin{equation}\label{eq:propu}
\begin{split}
\!\!\!\!\varphi \!=\!\varphi_1 \!\mathsf{U}_{(a,\kappa)}\! \varphi_2 \!\vee\! (\! \G_{(a,\kappa)} \!\varphi_1 \!\wedge\! \F\!_{\{\kappa\}} \!\!\left( \varphi_1 \!\wedge\! \varphi_2 \!\vee\! \varphi_1 \!\mathsf{U}_{(0,b-\kappa)}\! \varphi_2 )\! \right),
\end{split}
\end{equation}
where the temporal operators $\F$ and $\G$ in \eqref{eq:propu} have shorter intervals compared to the original interval $(\,a,\,b\,)$.
This form is referred to as \textit{syntactic separation} since it ensures syntactic equivalence~\cite{hunter2013expressive, bae2019bounded}, i.e., both sides have the same set of satisfying signals.

In this sense, we aim to decompose the overall synthesis problem into several subproblems with shorter horizons and fewer specifications, inspiring the modularized synthesis of the original specification. More precisely, we focus on the following fragment of complex formulas,
\begin{equation}\label{eq:stl_spec_vh}
\small \textstyle \varphi\!:=\! \underbrace{\textstyle \bigwedge_{i=1}^{n_s} \!\G_{[\,a^s_i,\,b^s_i\,]} \gamma^s_i}_{\mathsf{safety\ formula}} \!\wedge\! \underbrace{\textstyle \bigwedge_{j=1}^{n_p} \!\G_{[\,a^p_j,\,b^p_j\,]} \F_{[\,0,\,c^p_j\,]} \gamma^p_j}_{\mathrm{progress\ formula}} \!\wedge\! \underbrace{\textstyle \F_{[\,a^t,\,b^t\,]} \G_{[\,0,\,c^t\,]} \gamma^t}_{\mathrm{target\ formula}},
\end{equation}
where $\G_{[\,a^s_i,\,b^s_i\,]} \gamma^s_i$, $\G_{[\,a^p_j,\,b^p_j\,]} \F_{[\,0,\,c^p_j\,]} \gamma^p_j$, and $\F_{[\,a^t,\,b^t\,]} \G_{[\,0,\,c^t\,]} \gamma^t$ are the \textit{safety}, \textit{progress}, and \textit{target} subformulas, for $i\!\in\!\{1,2,\cdots,n_s\}$ and $j\!\in\!\{1,2,\cdots,n_p\}$, $n_s,n_p \!\in\! \mathbb{N}^+$, $\gamma^s_i$, $\gamma^p_j$, and $\gamma^t$ are the boolean formulas that only contain predicates connected with logical operators $\lnot$, $\wedge$, and $\vee$, and $[\,a^s_i,\,b^s_i\,]$, $[\,a^p_j,\,b^p_j\,]$, and $[\,a^t,\,b^t\,]$ are the non-empty \textit{syntactic intervals} of the subformulas. We also refer to $[\,a^s_i,\,b^s_i\,]$, $[\,a^p_j,\,b^p_j\!+\!c^p_j\,]$, and $[\,a^t,\,b^t\!+\!c^t\,]$ which represent the complete coverage of the subformulas as their \textit{complete intervals}, making a clear distinguishment with the \textit{syntactic intervals}. 

Similar to the popularly used GR(1) specifications~\cite{schlaipfer2011generalized}, the fragment $\varphi$ defined above is a complex STL formula composed of three components representing meaningful specifications for practical tasks. The \textit{safety} part consists of a series of \textit{always} subformulas specifying the conditions that should always hold. This could include for example safety rules applicable to the system. The \textit{progress} component contains the always subformulas with eventual operators embedded to represent the tasks that should be performed regularly, such as the monitoring routines. The \textit{target} component describes the task that should be achieved within a strict deadline. An always formula is embedded to ensure the holding of the target condition for a minimum time. 

Specifications in the form of eq. \eqref{eq:stl_spec_vh} already have a natural division in subformulas for individual subtasks. However, modularized synthesis also requires the division of the subformulas in time. 
Given a set of ordered timing points $\kappa_1\!<\!\kappa_2<\!\cdots\!<\!\kappa_l$, $\kappa_z\!\in\!\mathbb{N}$, $z\!\in\!\{1,2,\cdots,l\}$, $l\!\in\!\mathbb{N}$, we intend to split the specification $\varphi$ into subformulas $\varphi_z$ with shorter lengths. 
Then, modularized synthesis expects to efficiently solve the synthesis problem for $\varphi$ through a sequence of synthesis problems for its subformulas $\varphi_z$. 
To achieve this, the overlapping between the timing intervals of the subformulas should be eliminated to decouple the dependence of different timings. In the next section, we will show that the decoupling of the safety subformulas can be achieved by syntactic separation while ensuring the syntactic equivalence between the separated specification and the original one. However, the progress and the target subformulas are challenging to decouple since the timing overlapping cannot be eliminated by merely using syntactic separation. This has not been well investigated by existing work. We will also show that these subformulas can be decoupled using the complete specification split which ensures soundness but introduces conservativeness. Our work is the first to decompose such STL formulas for modularized synthesis in the literature.

\section{Main Results}\label{sec:main}

In this section, we will first show how to split the syntactic intervals of a complex specification while ensuring the syntactic equivalence. Then, we further present a complete split form to eliminate the overlapping of the complete intervals of the subformulas while causing certain conservativeness. Finally, we give the modularized synthesis algorithm based on these separation forms.

\subsection{Syntactic Timing Separation}\label{sec:synt}

The syntactic separation form of a complex specification in eq. \eqref{eq:stl_spec_vh} is defined as follows.

\begin{definition}[Syntactic Timing Separation]\label{def:s3}
 Given an ordered sequence of timing $0\!=\!\kappa_0\!<\!\kappa_1\!<\!\cdots\!<\!\kappa_l\!=\!L$, $\kappa_z\!\in\!\mathbb{N}$, for $z\!\in\!\{1,2,\cdots,l-1\}$, $l\!\in\!\mathbb{N}$, the specification $\varphi$ defined in eq.~\eqref{eq:stl_spec_vh} is said to be in a syntactic timing separation form if
\begin{equation}\label{eq:spl}
\varphi \textstyle := \bigwedge_{z=1}^l  \phi_z \wedge  \bigvee_{z=1}^l \phi^t_{z},
\end{equation}
where, for each $z\!\in\!\{1,2,\cdots,l\,\}$, 
\begin{equation*}
\begin{split}
&\small \phi_z \textstyle = \underbrace{\textstyle\bigwedge_{i=1}^{n_{s,z}} \G_{[\,a^s_{z,i},\,b^s_{z,i}\,]} \gamma^s_i}_{\mathsf{safety\  subformula}} \wedge \underbrace{\textstyle\bigwedge_{j=1}^{n_{p,z}} \G_{[\,a^p_{z,j},\,b^p_{z,j}\,]} \F_{[\,0,\,c^p_{z,j}\,]} \gamma^p_j}_{\mathrm{progress\  subformula}},\\
 & \small  \phi^t_{z} =\underbrace{\F_{[\,a^t_{z},\,b^t_{z}\,]} \G_{[\,0,\,c^t_{z}\,]} \gamma^t }_{\mathrm{target\ subformula}} \mbox{ or }    \phi^t_{z} =\neg\top,
\end{split}
\end{equation*}
and all intervals associated to $z\!\in\!\{1,2,\cdots,l\,\}$ satisfy $[\,a^s_{z,i},\,b^s_{z,i}\,]\!\subset\! [\,\kappa_{z-1},\,\kappa_{z}\,]$, $\forall\,i\in\{1,2,\cdots,n_{s,z}\}$, $[\,a^p_{z,j},\,b^p_{z,j}\,]\!\subset\!  [\,\kappa_{z-1},\,\kappa_{z}\,]$, $\forall\,j\!\in\!\{1,2,\cdots,n_{p,z}\}$, and
$[\,a^t_{z},\,b^t_{z}\,]\!\subset\!  [\,\kappa_{z-1},\, \kappa_{z}\,]$, where $n_{s,z}, n_{p,z}\!\in\!\mathbb{N}$ are the numbers of the effective safety and progress subformulas for $[\,\kappa_{z-1},\, \kappa_{z}\,]$.
$\qed$
\end{definition}

Consider the following specifications with the same length $6$: $\varphi_1 \!=\! \G_{[\,0,\,4\,]} \gamma_0 \!\wedge\! \G_{[\,2,\,6\,]} \gamma_1$, $\varphi_2 \!=\! \G_{[\,0,\,2\,]} \F_{[\,0,\,1\,]} \gamma_0 \!\wedge\! \G_{[\,2,\,6\,]} \gamma_1$, $\varphi_3\!=$ $\G_{[\,0,\,4\,]} \gamma_0 \!\wedge\! \F_{[\,0,\,4\,]}\! \G_{[\,0,\,2\,]} \gamma_1$, and $\varphi_4 \!=\! \G_{[\,0,\,2\,]} \!\F_{[\,0,\,1\,]} \gamma_0 \!\wedge\!\F_{[\,3,\,5\,]}\! \G_{[\,1,\,1\,]} \gamma_1$, 
where $\gamma_0$, $\gamma_1$ are boolean formulas. Consider splitting points $\kappa_0\!=\!0$, $\kappa_1\!=\!2$, $\kappa_2\!=\!6$, according to Definition~\ref{def:s3}, only $\varphi_2$ and $\varphi_4$ are in a form that is syntactically separated by $\kappa_0$, $\kappa_1$, $\kappa_2$. Formulas $\varphi_1$, $\varphi_3$ are not since $\kappa_1$ splits the intervals $[\,0,\,4\,]$.

We are interested in translating the specification $\varphi$ \eqref{eq:stl_spec_vh} into the separated form of \eqref{eq:spl} using syntactic separation. In this way, the individual numbers of the safety and progress specifications for each timing interval $z\in\{1,2,\cdots,l\}$, denoted as $n_{s,z}$, $n_{p,z}$, can be substantially smaller than those of the corresponding safety and progress specifications, i.e., $n_s$, $n_p$. The following syntactic separation principles can be used to transform \eqref{eq:stl_spec_vh} into \eqref{eq:spl} with syntactic equivalence guaranteed.

\begin{lemma}[\cite{hunter2013expressive}]\label{lm:ppfa}
The following properties hold for arbitrary STL formulas $\varphi$, $\varphi_1$, and $\varphi_2$, with $\kappa\!\in\!\mathbb{N}$, $\kappa \!<\!a$: $\F_{\{\kappa\}} (\lnot \varphi) \!=\! \lnot \F_{\{\kappa\}} \varphi$, $\F_{\{\kappa\}}(\varphi_1 \!\wedge\! \varphi_2)  \!=\! \F_{\{\kappa\}} \varphi_1 \!\wedge\! \F_{\{\kappa\}} \varphi_2$, $\F_{\{\kappa\}} (\varphi_1 \mathsf{U}_{(\,a,\,b\,)} \varphi_2) \!=\! \F_{\{\kappa\}} \varphi_1 \mathsf{U}_{(\,a,\,b\,)} \F_{\{\kappa\}} \varphi_2$, $\varphi \mathsf{U}_{(\,a,\,b\,)} (\varphi_1 \!\vee\! \varphi_2)=$ $ \varphi \mathsf{U}_{(\,a,\,b\,)} \varphi_1\!\vee\!\varphi \mathsf{U}_{(\,a,\,b\,)} \varphi_2$.
$\qed$
\end{lemma}

\begin{lemma}\label{lm:prop1}
The following conditions hold for arbitrary STL formulas $\varphi$, $\varphi_1$, and $\varphi_2$ defined in Sec.~\ref{sec:stl}.

1). $\F_{\{\kappa\}} \varphi = \G_{\{\kappa\}} \varphi$, for any $\kappa \in \mathbb{N}$, where both sides are true for signal $\xseq$, if and only if $(\xseq, \kappa) \vDash \varphi$.

2). $\G_{\{\kappa\}}(\varphi_1 \!\vee\! \varphi_2) \!=\! \G_{\{\kappa\}} \varphi_1 \!\vee\! \G_{\{\kappa\}} \varphi_2$ holds for any $\kappa \in \mathbb{N}$.

3). For any $a, b \!\in\! \mathbb{N}$, $a \leq b$, $\G_{\{\kappa\}} \!\!\left( \G_{[\,a,\,b\,]} \varphi \right) \!=\! \G_{[\,\kappa+a,\,\kappa+b\,]} \!\varphi$ and $\F_{\{\kappa\}} \!\left( \F_{[\,a,\,b\,]} \varphi \right) \!=\! \F_{[\,\kappa+a,\,\kappa+b\,]} \varphi$ hold for $\kappa \!\in\! \mathbb{N}$, $\kappa \!< \!a$.
$\qed$
\end{lemma}

\begin{lemma}\label{lm:prop15}
Given $a, b \in \mathbb{N}$, $a \leq b$ and an arbitrary STL formula $\varphi$, $\G_{[\,a,\,b\,]} \varphi = \G_{\{a\}} \varphi \wedge \G_{(\,a,\,b\,)} \varphi \wedge \G_{\{b\}} \varphi$ and $\F_{[\,a,\,b\,]} \varphi = \F_{\{a\}} \varphi \vee \F_{(\,a,\,b\,)} \varphi \vee \F_{\{b\}} \varphi$ hold.
$\qed$
\end{lemma}

\begin{theorem}[Arbitrary Syntactic Separation]\label{th:sepa}
Given {\small $\kappa \!\in\! \mathbb{N}$}, the following conditions hold for an STL formula $\varphi$,
\begin{equation}\label{eq:Ycom2}
\G_{[\,a,\,b\,]} \varphi \!=\! \G_{[\,a,\,\kappa\,]} \varphi \!\wedge\! \G_{[\,\kappa,\,b\,]} \varphi, ~
\F_{[\,a,\,b\,]} \varphi \!=\! \F_{[\,a,\,\kappa\,]} \varphi \!\vee\! \F_{[\,\kappa,\,b\,]} \varphi,
\end{equation}
with $a \!\leq\! \kappa \!\leq \!b$.
Moreover, the following conditions hold,
\begin{equation}\label{eq:arbspli}
\textstyle \G_{[\,\kappa_0,\,\kappa_l\,]} \varphi \!=\! \bigwedge_{i=1}^l \!\G_{[\,\kappa_{i-1},\,\kappa_i\,]} \varphi, ~ \F_{[\,\kappa_0,\,\kappa_l\,]} \varphi \!=\! \bigvee_{i=1}^l \!\F_{[\,\kappa_{i-1},\,\kappa_i\,]} \varphi,
\end{equation}
for $\kappa_0$, $\kappa_1$, $\cdots$, $\kappa_l \in \mathbb{N}$, $\kappa_0 < \kappa_1 < \cdots < \kappa_l$.
$\qed$
\end{theorem}

Lemmas~\ref{lm:prop1} and \ref{lm:prop15} provide complementary properties to previous work on syntactic separation~\cite{hunter2013expressive, bae2019bounded}. Note that they apply to all STL formulas as introduced in Sec.~\ref{sec:stl}, but not only the fragment in eq.~\eqref{eq:stl_spec_vh}. Most important is theorem \ref{th:sepa} which allows separating a subformula into the logical combination of shorter subformulas with an arbitrary number of timing points. Such separation as eq.~\eqref{eq:spl} does not change the syntax of the specification. i.e., for any signal $\xseq_{[\,0,\,L\,]}$ with $L\!=\!\mathcal{L}(\varphi)$, $\xseq \!\vDash \! \varphi \!\leftrightarrow\! \xseq \!\vDash \!\bigwedge_{z=1}^l  \phi_z \wedge  \bigvee_{z=1}^l \phi^t_{z}$. Nevertheless, syntactic timing separation only splits up the syntactic interval of a subformula, which does not eliminate the overlapping between the complete intervals of the subformulas. This is not sufficient for the modularized synthesis of specifications. In the following, we will give an alternative sufficient form of separation that is no longer equivalent to the original specification but ensures the separation of the complete intervals of the subformulas. This form is more conservative than the original specification but ensures the soundness of the solution and allows for modularized solutions to the synthesis problem.

\subsection{Complete Specification Split for Modularized Checking}\label{sec:msaf}

Before we give the complete splitting form of specification \eqref{eq:stl_spec_vh}, we first explain why eliminating the overlapping of the complete intervals of subformulas is important to modularized model checking which is the foundation of modularized synthesis to be explained in the following. Consider an STL specification $\varphi$ and a signal prefix $\xseq$ with the same length. For a given series of timing points $\kappa_0, \kappa_1, \cdots, \kappa_l$, modularized model checking investigates under what conditions and what subformulas $\tilde{\varphi}_1$, $\tilde{\varphi}_2$, $\cdots$, $\tilde{\varphi}_l$, where $\mathcal{L}(\bar{\varphi}_z) \!=\!\kappa_{z}\!-\!\kappa_{z-1}$ for all $z\!\in\!\{1,2,\cdots,l\}$, it ensures that
\begin{equation}\label{eq:split_verification}
\xseq_{[\,\kappa_{z-1},\kappa_{z}\,]}\vDash\bar{\varphi}_z,~\mathrm{for~some}~z\!\in\! \{1,2,\cdots,l\}
\rightarrow
\xseq \vDash \varphi.
\end{equation}
In such a way, we can split the model checking of the original signal $\xseq$ and specification $\varphi$ into $l$-steps of model checking for shorter signals $\xseq_{[\,\kappa_{z-1},\kappa_{z}\,]}\vDash\bar{\varphi}_z$ and specifications $\bar{\varphi}_z$. This is only feasible when the coverage or the complete interval of the subformulas $\bar{\varphi}_z$ is confined within the corresponding interval $[\,\kappa_{z-1},\kappa_{z}\,]$ such that it does not overlap with those of other subformulas. Otherwise, the model checking for the left side of \eqref{eq:split_verification} can not be performed independently for each $z\!\in\!\{1,2,\cdots,l\}$ due to the coupled timing dependence. 

Based on this consideration, we give the complete specification split form for formula $\varphi$ in eq.~\eqref{eq:stl_spec_vh} as
\begin{equation}\label{eq:complete_split}
\bar\varphi \textstyle:= \bigwedge_{z=1}^l  \bar\phi_z \wedge  \bigvee_{z=1}^l \bar\phi^t_z,
\end{equation}
where, for any $z\!\in\!\{1,2,\cdots,l\}$,
\begin{equation*}
\begin{split}
& \small \textstyle \bar{\phi}_z \!:=\! \bigwedge_{i=1}^{n_{s,z}} \!\G_{[\,a^s_{z,i},\,b^s_{z,i}\,]} \gamma^s_i \wedge 
	\bigwedge_{j=1}^{n_{p,z}} \G_{[\,a^p_{z,j},\,\min\{b^p_{z,j},\kappa_z-c^p_{z,j}\}\,]}  \F_{[\,0,\,c^p_{z,j}\,]} \gamma^p_j \\ 	& \textstyle~~~~ \wedge 
\bigwedge_{r=1}^{\hat{n}_{p,z}}\notag\!
	\F_{[\,\kappa_z-\tau_{z,r},\,\kappa_z\,]}\gamma^p_j
	\wedge 
	\bigwedge_{q=1}^{\hat{n}_{p,z-1}}
	\F_{[\,\kappa_{z-1},\,\kappa_{z-1}+c^p_{z-1,q}-\tau_{z-1,q}\,]}\gamma^p_j\\
&\small	\bar{\phi}^t_z \!: =\! \F_{[\,a^t_{z},\,\min\{b^t_{z},\, \kappa_z-c^t_z\}]} \G_{[\,0,\,{c}^t_{z}\,]} \gamma^t  \mbox{ or }    \bar{\phi}^t_z =\neg\top,
\end{split}
\end{equation*}
where $\hat{n}_{p,z}$ for any $z\!\in\!\{1,2,\cdots,l\}$, is the number of $j\!\in\!\{1,2,\cdots,n_{p,z}\}$ such that $b^p_{z,j} + c^p_{z,j} > \kappa_z$, i.e., the number of progress formulas that exceed the interval $[\,\kappa_{z-1},\,\kappa_z\,]$, and $\tau_{z,r}\!\in\![\,0,\,c^p_{z,r}\,]$ for any $r\!\in\!\{1,2,\cdots,\hat{n}_{p,z}\}$, is a heuristic value to be determined beforehand. 
It can be verified that $\mathcal{L}(\bar{\varphi})\!=\!\kappa_l\!=\!\mathcal{L}(\varphi)$. Then, we have the following two theorems to address the relation between the complete split form $\bar{\varphi}$ in eq.~\eqref{eq:complete_split} and the syntactic separation form $\varphi$ in eq.~\eqref{eq:spl}.
	
\begin{lemma}[Complete Interval Split]\label{lm:cis}
Given $0\!=\!\kappa_0\!<\!\kappa_1\!<\!\cdots\!<\!\kappa_l\!=\!L$, $\kappa_z\!\in\!\mathbb{N}$, $z\!\in\!\{1,2,\cdots,l-1\}$, $l\!\in\!\mathbb{N}$ and a specification $\bar{\varphi}$ in form \eqref{eq:complete_split}, if $\kappa_z-\kappa_{z-1} \geq c^{p}_{z,j}$ for all $j\!\in\!\{1,2,\cdots,n_{p,z}\}$ and for all $z\!\in\!\{1,2,\cdots,l\}$, the complete intervals of $\bar{\phi}_1$, $\bar{\phi}_2$, $\cdots$, $\bar{\phi}_l$ do not overlap, and the complete intervals $\bar{\phi}^t_1$, $\bar{\phi}^t_2$, $\cdots$, $\bar{\phi}^t_l$ do not overlap.
\end{lemma}

\begin{lemma}[Soundness]\label{lm:sound}
For a specification $\varphi$ in eq.~\eqref{eq:spl} and its complete split form $\bar{\varphi}$ in eq.~\eqref{eq:complete_split} with the splitting timing points $\kappa_0$, $\kappa_1$, $\cdots$, $\kappa_l$ as described in \textit{lemma~\ref{lm:cis}}, any signal prefix $\xseq$ with length $\mathcal{L}(\varphi)$ holds that $\xseq \vDash \bar{\varphi} \rightarrow \xseq \vDash \varphi$.
\end{lemma}

\begin{theorem}\label{th:sound}
For a specification $\varphi$ in eq.~\eqref{eq:spl} and its complete split form $\bar{\varphi}$ in eq.~\eqref{eq:complete_split} with the splitting timing points $\kappa_0$, $\kappa_1$, $\cdots$, $\kappa_l$ as described in \textit{lemma~\ref{lm:cis}}, $\xseq \vDash \varphi$ holds for signal $\xseq$ with length $\mathcal{L}(\varphi)$ if the following conditions both hold,\\
1). $\xseq_{[\,\kappa_{z-1},\kappa_{z}\,]}\vDash \G_{\{-\kappa_{z-1}\}} \bar{\phi}_z$, $\forall\,\,z\!\in\! \{1,2,\cdots,l\}$; \\
2). $\xseq_{[\,\kappa_{z-1},\kappa_{z}\,]}\vDash \G_{\{-\kappa_{z-1}\}} \bar{\phi}^t_z$, $\exists\,\,z\!\in\! \{1,2,\cdots,l\}$.
$\qed$
\end{theorem}

Theorem~\ref{th:sound} has solved the main problem of modularized model checking for specification $\varphi$ given in eq.~\eqref{eq:stl_spec_vh} by eliminating the overlapping between the complete intervals of its subformulas, as addressed by lemma~\ref{lm:cis}. From a practical perspective, the overlapping means that the timing coupling between different subtasks specifies that these subtasks need to be executed in parallel. In this sense, theorems \eqref{th:sound} provide a solution to decouple such dependence by imposing additional specifications to the subtasks, such that they can be solved independently in sequence. The soundness of the complete interval split is ensured by lemma~\ref{lm:sound}. The timing points that mark the solving sequence can be predetermined according to practical requirements.

\subsection{Modularized Synthesis of Split Specifications}

Given the complete specification split form $\bar{\varphi}$ in eq.~\eqref{eq:complete_split} for modularized model checking, we can further investigate the modularized synthesis by incorporating the constraints brought up by the dynamic systems \eqref{eq:opti_sys}. For this, we develop algorithm~\ref{ag:rme} for modularized synthesis of a split specification. In algorithm~\ref{ag:rme}, \textit{opt()} is a function of the optimization problem in eq.~\eqref{eq:opti_sto} with interface $(x_0,\,L,\,\varphi)$, and FEASIBLE is a binary variable to indicate whether problem~\eqref{eq:opti_sto} is feasible. Algorithm~\ref{ag:rme} allows us to perform synthesis for the dynamic system \eqref{eq:opti_sys} with specification $\bar{\varphi}$ in a modularized way, i.e., by solving a sequence of smaller synthesis problems in a timing order $\kappa_1$, $\kappa_2$, $\cdots$, $\kappa_l$. For each time $\kappa_z$, $z\!\in\!\{1,2,\cdots,l\}$, the synthesis subproblem requires substantially fewer integers than the original problem since it involves much shorter and fewer specifications.

\textbf{Complexity Analysis:} As addressed in Sec.~\ref{sec:prob_state}, the complexity of directly synthesizing the original specification $\varphi$ in eq.~\ref{eq:stl_spec_vh} is $O(2^{NL})$, where $L\!=\!\mathcal{L}(\varphi)$ is the length of $\varphi$ and $N:=\max_{x\in\{1,2,\cdot,l\}}(n_s+n_p+1)$ is the total number of subformulas. For its complete split form $\bar{\varphi}$ in eq.~\eqref{eq:complete_split}, assume that the longest subformula has a length $\bar{L}\!=\!\max_{z\in\{1,2,\cdots,l\}}(\kappa_z-\kappa_{z-1})$, the complexity of algorithm~\ref{ag:rme} is $O(l\cdot2^{\bar{N}\bar{L}})$, where $\bar{N}\!:=\!\max_{z\in\{1,2,\cdots,l\}}(n_{s,z} \!+\! n_{p,z}\!+\!\hat{n}_{p,z}\!+\!\hat{n}_{p,z-1})$ denotes the maximum number of subformulas in one synthesis module $z\in\{1,2,\cdots,l\}$. As addressed in Sec.~\ref{sec:synt} and Sec.~\ref{sec:msaf}, from syntactic separation we can expect $n_{s,z} \ll n_s$ and $\hat{n}_{p,z} < n_{p,z} \ll n_p$ for any $z\!\in\{1,2,\cdots,l\}$, which leads to $\bar{N} \ll N$. Moreover, we can also ensure $\bar{L} \ll L$ by properly selecting the timing points $\kappa_0$, $\kappa_1$, $\cdots$, $\kappa_l$. Thus, with $2^{\bar{N}\bar{L}} \ll 2^{NL}$, modularized synthesis can substantially reduce the complexity of the synthesis problem for long and complex specifications and improve its efficiency.

\begin{algorithm}[htbp]
\caption{Modularized Synthesis of Specification $\bar{\varphi}$}
\label{ag:rme}
\begin{algorithmic}[1]
\renewcommand{\algorithmicrequire}{\textbf{Input:}}
\renewcommand{\algorithmicensure}{\textbf{Output:}}
\REQUIRE Initial system condition $x_0$, $\kappa_0\!=\!0$, splitting timing points $\kappa_z$ and subformulas $\bar{\phi}_z$, $\bar{\phi}^t_z$, for $z\!\in\!\{1,2,\cdots,l\}$.
\ENSURE Control signal $\useq$ and state signal $\xseq$.
\STATE {$x_{\kappa_0} \leftarrow x_0$}
\FOR {$z=1$ \TO $l$} 
	\STATE {$L_z \leftarrow \kappa_z-\kappa_{z-1}$}
	\IF  {$z>1$ and $\xseq_{[\,\kappa_0,\kappa_{z-1}\,]} \!\vDash\! \vee_{w}^{z-1} \bar{\phi}^t_w$}
		\STATE {$\xseq_{[\,\kappa_{z-1}+1,\,\kappa_z\,]},\, \useq_{[\,\kappa_{z-1},\,\kappa_z-1\,]} \leftarrow \mathrm{opt}(x_{\kappa_{z-1}},\, L_z,\, \bar{\phi}_z)$}
	\ELSE
		\STATE {$\xseq_{[\,\kappa_{z-1}+1,\,\kappa_z\,]},\, \useq_{[\,\kappa_{z-1},\,\kappa_z-1\,]} \leftarrow \mathrm{opt}(x_{\kappa_{z-1}},\, L_z,\, \bar{\phi}_z \wedge\bar{\phi}^t_z)$}
		\IF {\NOT FEASIBLE}		
			\STATE $\xseq_{[\,\kappa_{z-1}+1,\,\kappa_z\,]},\, \useq_{[\,\kappa_{z-1},\,\kappa_z-1\,]} \leftarrow \mathrm{opt}(x_{\kappa_{z-1}},\, L_z,\, \bar{\phi}_z)$
		\ENDIF
	\ENDIF
\ENDFOR
\STATE $\useq \leftarrow \useq_{[\,\kappa_0,\,\kappa_l-1\,]}$, $\xseq \leftarrow \xseq_{[\,\kappa_0,\,\kappa_l\,]}$
\end{algorithmic}
\end{algorithm}

\textbf{Limitations:} Nevertheless, a limitation of algorithm~\ref{ag:rme} is that it only ensures soundness but not optimality nor completeness to the original problem. This means that if it generates a feasible solution $\xseq$, it is certainly a feasible solution to the synthesis problem of the original specification, i.e., $\xseq \!\vDash \!\varphi$ (soundness). However, it might not be the optimal solution in terms of the robustness $\varrho(\xseq, \varphi)$, i.e., local optimality does not necessarily lead to global optimality. Moreover, if algorithm~\ref{ag:rme} is not feasible, it does not mean that the original synthesis problem $\xseq \!\vDash \!\varphi$ is also infeasible (completeness). However, this is already sufficient for most practical robotic tasks. 
It is also worth noting that algorithm~\ref{ag:rme} might not be feasible for an arbitrary initial system condition $x_0$. For a system~\eqref{eq:opti_sto} and a specification $\bar{\varphi}$ in eq.~\eqref{eq:complete_split}, the initial condition $x_0$ that ensures the feasibility of $\xseq\!\vDash\!\varphi$ belongs to a set which is referred to as the \textit{largest satisfaction region}~\cite{belta2017largest}. How to utilize the feasible sets to improve the feasibility of a synthesis problem is also partially discussed in a recent work~\cite{yu2022model}. We are not providing further discussions on this since it is out of the scope of this paper.

\section{Case Study in Simulation}\label{sec:casestudy}

\begin{figure}[htbp]
\centering
\includegraphics[width=0.4\textwidth]{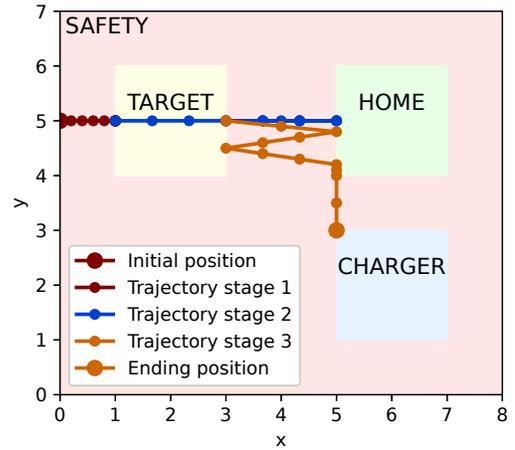}
\caption{The illustration of the robot monitoring scenario with the spatial information of the synthesized trajectory subject to specification $\bar{\varphi}$.}
\label{fig:region}
\end{figure}

In this section, we use an essential simulation study to showcase how the proposed modularized synthesis approach can be used to efficiently solve a synthesis problem for a complex specification. As shown in Fig.~\ref{fig:region}, we consider a scenario where a mobile robot is required to perform a monitoring task in a rectangular space SAFETY sized $8\,\times 7\,$ (red) with three square regions TARGET (yellow), HOME (green), and CHANGER (blue) which are centered at $(2, 5)$, $(6, 5)$, and $(6, 2)$ with the same side length $2$. The robot is described as the following dynamic model,
\begin{equation}\label{eq:sing_int}
\zeta_{k+1} = \zeta_k + u_k,~k\!\in\! \mathbb{N},
\end{equation}
where $\zeta_k\!\in\! \mathbb{R}^2$ denotes the planar coordinate of the robot position at time step $k$ and $u_k \!\in\! \mathbb{R}^2$ is the position increments of the robot per step as the control input of the system. The control input of the system is subject to saturation constraints $\left| u_{k,1} \right| \!\leq\! 1$, $\left| u_{k,2} \right| \!\leq\! 1$ for all $k\!\in\!\mathbb{N}$, where $u_{k,1}, u_{k,2} \!\in\! \mathbb{R}$ are the elements of $u_k$. The monitoring task is described as follows.

1). Starting from position $(0,5)$, the robot should frequently visit TARGET every $5$ steps or fewer until $k\!=\!40$.

2). From $k\!=\!15$ to $k\!=\!45$, once the robot leaves HOME, it should get back to HOME within $5$ time steps.

3). After $k\!=\!20$ and before $k\!=\!45$, it should stay in CHANGER continuously for at least $3$ time steps to charge.

4). The robot should always stay in the SAFETY region. 

These tasks can be specified using the following formulas: $\G_{[\,0,\,35\,]} \F_{[\,0,\,5\,]}\!\gamma_T$, $\G_{[\,15,\,40\,]}  \F_{[\,0,\,5\,]} \gamma_H$, $\F_{[\,20,\,42\,]} \G_{[\,0,\,3\,]} \gamma_C$, and $\G_{[\,0,\,45\,]} \gamma_S$ respectively, where $\gamma_T$, $\gamma_H$, $\gamma_C$, and $\gamma_S$ are boolean formulas used to specify $\xi_k \!\in$TARGET, $\xi_k \!\in$HOME, $\xi_k \!\in$CHARGER, and $\xi_k\!\in$SAFETY, for $k\!\in\!\mathbb{N}$. Thus, the overall robot task is the conjunction of these formulas. With splitting timing points $\kappa_0\!=\!0$, $\kappa_1\!=\!15$, $\kappa_2\!=\!30$, and $\kappa_3\!=\!45$, the overall specification can be represented as a syntactic separation form as eq.~\eqref{eq:spl}, i.e., $\varphi \!:=\! \bigwedge_{z=1}^3 \phi_z \!\wedge\! \bigwedge_{z=1}^3 \phi^t_z$, where $\phi_1\!=\!\G_{[\,0,\,15\,]} \gamma_S \wedge \G_{[\,0,\,15\,]}\F_{[\,0,\,5\,]} \gamma_T$, $\phi_2\!=\!\G_{[\,15,\,30\,]} \gamma_S \!\wedge\! \G_{[\,15,\,30\,]}\F_{[\,0,\,5\,]} \gamma_T \!\wedge\! \G_{[15,30]}\F_{[0,5]}\! \gamma_H$, 
$\phi_3\!=\!\G_{[30,45]} \!\gamma_S \!\wedge\! \G_{[30,35]}\F_{[0,5]} \!\gamma_T \!\wedge\! \G_{[30,40]}\F_{[0,5]} \!\gamma_H$, 
$\phi^t_1\!=\!\lnot \top$, 
$\phi^t_2\!=\!\F_{[\,20,\,30\,]}\G_{[\,0,\,3\,]} \gamma_C$, 
$\phi^t_3\!=\!\F_{[\,30,\,42\,]}\G_{[\,0,\,3\,]} \gamma_C$.

\begin{figure}[htbp]
\centering
\includegraphics[width=0.4\textwidth]{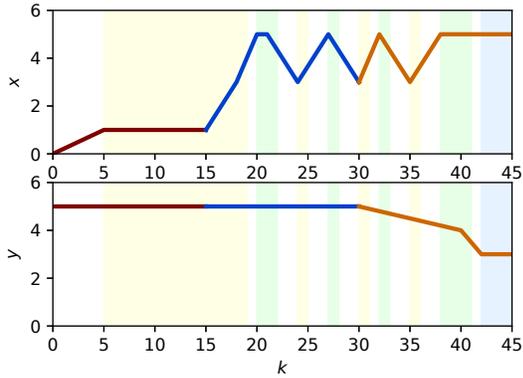}
\caption{The $x$- and $y$-positions of the robot trajectory in three stages. The color in the background indicates in which region the robot stays, namely yellow for TARGET, green for HOME, and blue for CHARGER, which is consistent with Fig.~\ref{fig:region}.}
\label{fig:traj}
\end{figure}

We transform the overal specification $\varphi$ into a complete split form as eq.~\eqref{eq:complete_split}, i.e., $\bar{\varphi} := \bigwedge_{z=1}^3 \bar{\phi}_z \!\wedge\! \bigwedge_{z=1}^3 \bar{\phi}^t_z$, where
$\bar{\phi}_1\!=\! \G_{[\,0,\,15\,]} \gamma_S \!\wedge\! \G_{[\,0,\,10\,]}\F_{[\,0,\,5\,]} \gamma_T \!\wedge\! \F_{[\,12,\,15\,]} \gamma_T$,
$\bar{\phi}_2\!=\! \G_{[\,15,\,30\,]} \gamma_S \!\wedge\! \F_{[\,15,\,17\,]} \gamma_T \!\wedge\G_{[\,15,\,25\,]}\F_{[\,0,\,5\,]} \gamma_T \!\wedge\! \F_{[\,27,\,30\,]} \gamma_T \!\wedge\! \G_{[\,15,\,25\,]}\F_{[\,0,\,5\,]} \gamma_H \!\wedge\! \F_{[\,27,\,30\,]} \gamma_H$,
$\bar{\phi}_3\!= \G_{[\,30,\,45\,]} \gamma_S \!\wedge\! \!\F_{[\,30,\,32\,]}\gamma_T \!\wedge\! 
\G_{[\,30,\,35\,]}\F_{[\,0,\,5\,]} \gamma_T \!\wedge\! 
\F_{[\,30,\,32\,]}\gamma_H \!\wedge\! \G_{[\,30,\,40\,]}\F_{[\,0,\,5\,]} \gamma_H$,
$\bar{\phi}^t_1\!=\!\lnot \top$,
$\bar{\phi}^t_2\!=\!\F_{[\,20,\,27\,]}\G_{[\,0,\,3\,]} \gamma_S$,
$\bar{\phi}^t_3\!=\!\F_{[\,30,\,42\,]}\G_{[\,0,\,3\,]} \gamma_S$, where the heuristic $\tau$ values are all determined as $3$.
Then, we use Algorithm~\ref{ag:rme} to solve an open-loop control signal for system \eqref{eq:sing_int} with the split specification $\bar{\varphi}$. 
The \textit{stlpy} toolbox~\cite{kurtz2022mixed} is used to implement the \textit{opt()} method in algorithm~\ref{ag:rme}. The program for this simulation study is published at~\cite{zhang2023code}. The resulting robot trajectory $\zeta_k$ is shown in Fig.~\ref{fig:region} and Fig.~\ref{fig:traj}. The trajectories in different stages are marked with different colors. 

From Fig.~\ref{fig:region} and Fig.~\ref{fig:traj}, it can be seen that the robot starts from the initial position $(0,5)$, reaches the TARGET at $k\!=\!5$ and stays there until $k\!=\!15$. From $k\!=\!15$, the robot oscillates between TARGET and HOME to satisfy the task requirements 1) and 2). After $k\!=\!40$, the robot maintains the visiting frequency to HOME, while taking time to charge itself, which satisfies condition 3). During the entire process, the robot is restricted within the SAFE region, which satisfies specification 4). Therefore, all task specifications are satisfied, which indicates the efficacy of the proposed timing separation approaches and modularized synthesis methods.

\section{CONCLUSIONS}\label{sec:conclu}

In this paper, we discuss how to split a big synthesis problem for a complex and long STL specification into several smaller optimization problems with less complexity. The two proposed separation forms for the specification, namely a syntactically separated form and a complete splitting form, allow us to solve these smaller problems in a modularized manner, which is an important step toward efficient optimization-based specification synthesis. There are still two limitations of our work. One is that we only investigate the modularized synthesis for a certain class of STL formulas, although it is sufficient for many practical tasks. The other one is that the feasibility condition of modularized synthesis has not been deeply studied. Our future work will extend the results to wider fragments of STL formulas. We will also incorporate the feasible sets of specifications to investigate the feasibility of modularized synthesis.




\section*{APPENDIX}\label{sec:app}

\subsection{Proof of Lemma~\ref{lm:prop1}}

For 1), from the definition of STL syntax, we know, for an arbitrary STL formula $\varphi$, $\F_{\{\kappa\}} \varphi = \top \mathsf{U}_{\{\kappa\}} \varphi$ indicates that there must exist $k \in \{\kappa\}$, such that $(x,k) \vDash \varphi$, which means $(x,\kappa) \vDash \varphi$. Then, from $\G_{\{\kappa\}} \varphi = \lnot \F_{\{\kappa\}} \lnot \varphi$ and lemma~\ref{lm:ppfa}, we know $\G_{\{\kappa\}} \varphi \leftrightarrow \lnot \left( \lnot \F_{\{\kappa\}} \varphi \right) = \F_{\{\kappa\}} \varphi$.

For 2), according to lemma~\ref{lm:ppfa}, we have $\F_{\{\kappa\}}(\varphi_1 \!\vee\! \varphi_2) = \lnot \F_{\{\kappa\}} (\lnot \varphi_1 \!\wedge\! \lnot \varphi_2) \!=\! \lnot (\F_{\{\kappa\}} \lnot \varphi_1 \wedge \F_{\{\kappa\}}  \lnot \varphi_2)$.
Then, using lemma~\ref{lm:ppfa}, we obtain $\F_{\{\kappa\}}(\varphi_1 \!\vee\! \varphi_2) = \lnot (\lnot \F_{\{\kappa\}} \varphi_1 \!\wedge\! \lnot \F_{\{\kappa\}} \varphi_2) = \F_{\{\kappa\}} \varphi_1 \vee \F_{\{\kappa\}} \varphi_2$,
which also holds for $\G$, i.e., $\G_{\{\kappa\}}(\varphi_1 \vee \varphi_2) = \G_{\{\kappa\}} \varphi_1 \vee \G_{\{\kappa\}} \varphi_2$, according to 1) of this lemma.

For 3), for $\kappa < a$, we have
\begin{equation}\label{eq:pf1}
\F_{\{\kappa\}} \!\left( \F_{[\,a,\,b\,]} \varphi \right)= \top \mathsf{U}_{[\,\kappa+a,\,\kappa+b\,]} \varphi = \F_{[\,\kappa+a,\,\kappa+b\,]} \varphi.
\end{equation}
Applying lemma~\ref{lm:ppfa} to $\F_{\{\kappa\}} \G_{(\,a,\,b\,)} \varphi$, we have $\F_{\{\kappa\}} \G_{(\,a,\,b\,)} \varphi = \F_{\{\kappa\}} \left( \lnot \F_{(\,a,\,b\,)} \lnot \varphi \right) = \lnot \F_{\{\kappa\}} \!\left( \F_{(\,a,\,b\,)} \lnot \varphi \right)$.
Considering \eqref{eq:pf1}, we have $\F_{\{\kappa\}} \G_{(\,a,\,b\,)} \varphi = \lnot \F_{(\,\kappa+a,\,\kappa+b\,)} \lnot \varphi = \G_{(\,\kappa+a,\,\kappa+b\,)} \varphi$.
According to condition 1), we know,
\begin{equation}\label{eq:pf2}
\G_{\{\kappa\}} \G_{(\,a,\,b\,)} \varphi = \F_{\{\kappa\}} \G_{(\,a,\,b\,)} \varphi = \G_{(\,\kappa+a,\,\kappa+b\,)} \varphi.
\end{equation}
Therefore, principle 3) is proved by \eqref{eq:pf1} and \eqref{eq:pf2}.

\subsection{Proof of Lemma~\ref{lm:prop15}}

Applying eq.~\eqref{eq:propu} to formulate formula $\F_{[\,a,\,b\,]} \varphi$, we have
\begin{equation}\label{eq:Fdic}
\begin{split}
\F_{[\,a,\,b\,]} \varphi = \top \mathsf{U}_{[\,a,\,b\,]} \varphi \,&= \F_{\{a\}} \varphi \vee \top \mathsf{U}_{(\,a,\,b\,)} \varphi \vee \top \mathsf{U}_{\{b\}} \varphi \\
\,& = \F_{\{a\}} \varphi \vee \F_{(\,a,\,b\,)} \varphi \vee \F_{\{b\}} \varphi.
\end{split}
\end{equation}
Then, applying \eqref{eq:Fdic} to $\G_{[\,a,\,b\,]} \varphi$ we obtain
\begin{equation}\label{eq:Gdic}
\begin{split}
\G_{[a,b]} \varphi &=\! \lnot \F_{[a,b]} \lnot \varphi = \lnot\left( \F_{\{a\}} \lnot \varphi \!\vee\! \F_{(a,b)} \lnot \varphi \!\vee\! \F_{\{b\}} \lnot \varphi \right) \\
\,&= \lnot \F_{\{a\}} \lnot \varphi \wedge \lnot \F_{(\,a,\,b\,)} \lnot \varphi \wedge \lnot \F_{\{b\}} \lnot \varphi \\
\,&= \G_{\{a\}} \varphi \wedge \G_{(\,a,\,b\,)} \varphi \wedge \G_{\{b\}} \varphi.
\end{split}
\end{equation}
Thus, this lemma is proved by \eqref{eq:Fdic} and \eqref{eq:Gdic}.

\subsection{Proof of Theorem \ref{th:sepa}}

Substituting $\varphi_1 = \top$ and $\varphi_2 = \varphi$ to \eqref{eq:propu}, we have
\begin{equation*}
\begin{split}
\top \mathsf{U}_{(\,a,\,b\,)} \varphi = & \,
\top \mathsf{U}_{(\,a,\,\kappa\,)} \varphi \vee ( \G_{(\,a,\,\kappa\,)} \!\top\! \wedge \F_{\{\kappa\}} \!\left(  \varphi \vee \!\top \mathsf{U}_{(\,0,\,b-\kappa\,)} \varphi ) \right), \\
=& \,\F_{(\,a,\,\kappa\,)} \varphi \vee \F_{\{\kappa\}} \!\left( \varphi \vee \F_{(\,0,\,b-\kappa\,)} \varphi \right).
\end{split}
\end{equation*}
Applying properties 1) and 3) in Lemma \ref{lm:prop1}, we obtain
\begin{equation*}
\begin{split}
\F_{(\,a,\,b\,)} \varphi \!=& \top \mathsf{U}_{(\,a,\,b\,)} \varphi\!=\! \F_{(\,a,\,\kappa\,)} \varphi \!\vee\! \F_{\{\kappa\}}  \varphi \!\vee\! \F_{\{\kappa\}} \F_{(\,0,\,b-\kappa\,)} \varphi \\
=& \F_{(\,a,\,\kappa\,)} \varphi \vee \F_{\{\kappa\}}  \varphi \vee \F_{(\,\kappa,\,b\,)} \varphi.
\end{split}
\end{equation*}
Substituting it to \eqref{eq:Fdic}, we obtain
\begin{equation*}
\begin{split}
\,&\F_{[\,a,\,b\,]} \varphi = \F_{(\,a,\,\kappa\,)} \varphi \vee \F_{\{\kappa\}}  \varphi \vee \F_{(\,\kappa,\,b\,)} \varphi \\
\,&~=\! \F_{\{a\}} \varphi \vee \F_{(\,a,\,\kappa\,)} \varphi \vee \F_{\{\kappa\}}  \varphi \vee \F_{(\,\kappa,\,b\,)} \varphi \vee \F_{\{b\}}  \varphi \\
\,&~=\! \F_{[a,\kappa]} \varphi \vee \!\left( \F_{\{\kappa\}}  \varphi \!\vee\! \F_{(\kappa,b)} \varphi \!\vee\! \F_{\{b\}} \right) \!=\! \F_{[a,\kappa]} \varphi \!\vee\! \F_{[\kappa,b]} \varphi.
\end{split}
\end{equation*}
Then, applying it to $\G_{[\,a,\,b\,]} \varphi$, we obtain
\begin{equation*}
\begin{split}
\G_{[\,a,\,b\,]}& \, \varphi =  \lnot \F_{(\,a,\,b\,)} \lnot \varphi = \lnot \left( \F_{[\,a,\,\kappa\,]} \lnot \varphi \vee \F_{[\,\kappa,\,b\,]} \lnot \varphi \right) \\
= &  \, \lnot \F_{[\,a,\,\kappa\,]} \lnot \varphi \wedge \lnot \F_{[\,\kappa,\,b\,]} \lnot \varphi = \G_{[\,a,\,\kappa\,]}  \varphi \wedge  \G_{[\,\kappa,\,b\,]}  \varphi.
\end{split}
\end{equation*}
The two equations above prove \eqref{eq:Ycom2}. Also, \eqref{eq:arbspli} can be proved by recursively applying \eqref{eq:Ycom2} to the time points $\kappa_0$, $\kappa_1$, $\cdots$, $\kappa_l$.

\subsection{Proof of Lemma~\ref{lm:cis}}

For any $z\!\in\!\{1,2,\cdots,l\}$, we know $[\,a^s_{z,i},\,a^s_{z,i}\,] \!\subset\! [\,\kappa_{z-1},\,\kappa_z\,]$ for all $i\!\in\!\{1,2,\cdots,n_{s,z}\}$ and $[\,a^p_{z,j},\,a^p_{z,j}\,] \!\subset\! [\,\kappa_{z-1},\,\kappa_z\,]$ for all $j\!\in\!\{1,2,\cdots,n_{p,z}\}$. Thus, given $\tau_{z,r}\!\in\![\,0,\,c^p_{z,r}\,]$ and $\kappa_{z}-\kappa_{z-1}\!\geq\!c^{p}_{z-1,j}$, the complete interval of $\bar{\phi}_z$ reads $[\,\min\{\kappa_{z-1}, a^s_{z,i}$, $a^s_{z,i}\},\,\max\{\kappa_z, b^s_{z,i},b^s_{z,i}\}\,] \!=\! [\,\kappa_{z-1},\,\kappa_z\,]$ since $\bar{\varphi}_z$ in eq.~\eqref{eq:complete_split} is already in a syntactic separation form. Thus, we know that the complete intervals of $\bar{\phi}_z$ for all $z\!\in\!\{1,2,\cdots,l\}$ do not overlap. Then, for the complete interval of $\bar{\phi}^t_z$, we have $[\,a^t_z,\, \min\{\kappa_z, b^t_z + c^t_z\}\,]\!\subset \! [\,\kappa_{z-1},\,\kappa_z\,]$, which implies that the complete intervals of $\bar{\phi}^t_z$ for all $z\!\in\!\{1,2,\cdots,l\}$ do not overlap.

\subsection{Proof of Lemma~\ref{lm:sound}}

It can be noticed that $\bar{\varphi}$ and $\varphi$ share the same safety formulas and also the same progress formulas with $b^p_{z,j} \!+\! c^p_{z,j} \!<\! \kappa_z$, for $j\!\in\!\{1,2,\cdots,n^p_{z,j}\}$, for all $z\!\in\!\{1,2,\cdots,l\}$. Thus, these subformulas already have split complete intervals and ensure the same semantics between $\bar{\varphi}$ and $\varphi$.  Thus, we directly look into the progress formulas with $b^p_{z,j} \!+\! c^p_{z,j}\! >\! \kappa_z$ for which we have $\G_{[\,a^p_{z,j},\,b^p_{z,j}\,]} \F_{[\,0,\,c^p_{z,j}\,]} \gamma^p_j = \G_{[\,a^p_{z,j},\,\kappa_z-c^p_{z,j}\,]} \F_{[\,0,\,c^p_{z,j}\,]} \gamma^p_j \wedge \G_{[\,\kappa_z-c^p_{z,j},\,b^p_{z,j}\,]} \F_{[\,0,\,c^p_{z,j}\,]} \gamma^p_j$ using theorem~\ref{th:sepa}. Note that $\G_{[\,\kappa_z-c^p_{z,j},\,b^p_{z,j}\,]} \F_{[\,0,\,c^p_{z,j}\,]} \gamma^p_j $ means that, there should always exist $\kappa_z\!-\!c^p_{z,j}\!<\!k_1\!<\!k_2\!<\!b^p_{z,j}\!+\!c^p_{z,j}$ and $k_2\!-\!k_1 \!<\! c^p_{z,j}$, such that $(\xseq, k_1)\vDash \gamma^p_j$ and $(\xseq, k_2)\vDash \gamma^p_j$. In this sense, it is straightforward to infer that, for any $\tau_{z,j}\!\in\![\,\kappa_z\!-\!b^p_{z,j},\,c^p_{z,j}\,]$, we have $\F_{[\,\kappa_z-\tau_{z,j},\,\kappa_z\,]}\gamma^p_j \!\wedge\! \F_{[\,\kappa_z,\,\kappa_z-\tau_{z,j}+c^p_{z,j}\,]}\gamma^p_j \!\rightarrow\! \G_{[\,\kappa_z-c^p_{z,j},\,b^p_{z,j}\,]} \F_{[\,0,\,c^p_{z,j}\,]} \gamma^p_j$. Therefore, we have $\bigwedge_{z=1}^l\!\bar{\phi}_z \!\rightarrow\! \bigwedge_{z=1}^l\!\phi_z$.

Now, looking into the target subformulas, it is easy to verify that $\F_{[\,a^t_{z},\,\min\{b^t_{z},\, \kappa_z-c^t_z\}]} \G_{[\,0,\,{c}^t_{z}\,]} \gamma^t \!\rightarrow \! \F_{[\,a^t_{z},\,b^t_{z}\,]} \G_{[\,0,\,{c}^t_{z}\,]} \gamma^t$ holds for all $z\!\in\!\{1,2,\cdots,l\}$, which leads to $\bigvee_{z=1}^l\!\bar{\phi}^t_z \!\rightarrow\! \bigvee_{z=1}^l\!\phi^t_z$. Then, we can summarize that $\bigwedge_{z=1}^l\!\bar{\phi}_z \!\wedge\! \bigvee_{z=1}^l\!\bar{\phi}^t_z \!\rightarrow\! \bigwedge_{z=1}^l\!\phi_z \!\wedge\! \bigvee_{z=1}^l\!\phi^t_z$ which leads to $\xseq \vDash \bar{\varphi} \rightarrow \xseq \vDash \varphi$.

\subsection{Proof of Theorem~\ref{th:sound}}

We first look into condition 1). For $z\!=\!l$, it addresses $\xseq_{[\,\kappa_{l-1},\kappa_{l}\,]}\vDash \G_{\{-\kappa_{l-1}\}} \bar{\phi}_l$. Also, according to the semantics of STL formulas, for any $\xseq_{[\,\kappa_{z},\kappa_{l}\,]}\vDash \varphi'_z$, where $\varphi'_z$ is an STL formula, for $z\!\in\!\{1,2,\cdots,l-1\}$, if $\xseq_{[\,\kappa_{z-1},\kappa_{z}\,]}\vDash \G_{\{-\kappa_{z-1}\}} \bar{\phi}_z$ and the complete intervals of $\bar{\phi}_z$ and $\varphi'_z$ do not overlap, we have $\xseq_{[\,\kappa_{z-1},\kappa_{l}\,]} \vDash \G_{\{-\kappa_{z-1}\}} \bar{\phi}_z \wedge \G_{\{\kappa_{z}-\kappa_{z-1}\}} \varphi'_z$. Applying this recursively from $z\!=\!l$ back to $z\!=\!1$, we obtain $\xseq_{[\,\kappa_0,\,\kappa_l\,]} \vDash \bar{\phi}_1\wedge\bar{\phi}_2 \wedge \cdots \wedge \bar{\phi}_l$, i.e., $\xseq \vDash \bigwedge_{z=1}^l \bar{\phi}_z$. This inference indicates 1)$\,\rightarrow\!\xseq \vDash \bigwedge_{z=1}^l \bar{\phi}_z$. Note that this only holds if the complete intervals of $\bar{\phi}_1$, $\bar{\phi}_2$, $\cdots$, $\bar{\phi}_l$ do not overlap, which is ensured by lemma~\ref{lm:cis}.

For condition 2), if there exists $z\!\in\!\{1,2,\cdots,l\}$, such that $\xseq_{[\,\kappa_{z-1},\kappa_{z}\,]}\vDash \G_{\{-\kappa_{z-1}\}} \bar{\phi}^t_z$, we know that there exists $k\!\in\![\,a_z^t,\,\min\{b^t_{z},\, \kappa_z-c^t_z\}]$, such that $\xseq_{[\,k,\,k+c^t_z\,]} \vDash \G_{[\,0,\,{c}^t_{z}\,]} \gamma^t$ which implies $\xseq \vDash \bigvee_{z=1}^l \bar{\phi}^t_z$ since $[\,a_z^t,\,\min\{b^t_{z},\, \kappa_z-c^t_z\}] \subset [\,\kappa_0,\,\kappa_l\,]$ for all $z$. This renders 2)$\,\rightarrow\! \bigvee_{z=1}^l \bar{\phi}^t_z$. Therefore, we can summarize that 1) $\wedge$ 2) $\rightarrow\!\xseq \!\vDash \! \bigwedge_{z=1}^l \bar{\phi}_z \!\wedge\! \bigvee_{z=1}^l \bar{\phi}^t_z\!\leftrightarrow\!\xseq\!\vDash\!\bar{\varphi}$. Considering lemma~\ref{lm:sound}, we further have 1) $\wedge$ 2) $\rightarrow\!\xseq\!\vDash\!\varphi$.
\balance

\bibliographystyle{IEEEtran}
\bibliography{IEEEabrv, reference.bib}

\end{document}